\documentclass[runningheads]{llncs}
\usepackage[T1]{fontenc}
\usepackage{amsmath} 
\usepackage{amssymb}
\usepackage{graphicx}
\usepackage{xcolor}
\definecolor{mypink1}{rgb}{0.858, 0.188, 0.478}

\begin{document}

\title{Mamba-UNet: UNet-Like Pure Visual Mamba for Medical Image Segmentation}

\titlerunning{Mamba-UNet for Medical Image Segmentation}

\author{Ziyang Wang\inst{1}, Jian-Qing Zheng\inst{1}, Yichi Zhang\inst{2}, Ge Cui\inst{3}, Lei Li\inst{4} }

\institute{
$^1$ University of Oxford, UK \\
$^2$ Fudan University, China \\
$^3$ University of Pittsburgh, USA \\
$^4$ University of Copenhagen, DK \\
\email{ziyang.wang@cs.ox.ac.uk}}

\maketitle              
\begin{abstract}

In recent advancements in medical image analysis, Convolutional Neural Networks (CNN) and Vision Transformers (ViT) have set significant benchmarks. While the former excels in capturing local features through its convolution operations, the latter achieves remarkable global context understanding by leveraging self-attention mechanisms. However, both architectures exhibit limitations in efficiently modeling long-range dependencies within medical images, which is a critical aspect for precise segmentation. Inspired by the Mamba architecture, known for its proficiency in handling long sequences and global contextual information with enhanced computational efficiency as a State Space Model (SSM), we propose Mamba-UNet, a novel architecture that synergizes the U-Net in medical image segmentation with Mamba's capability. Mamba-UNet adopts a pure Visual Mamba (VMamba)-based encoder-decoder structure, infused with skip connections to preserve spatial information across different scales of the network. This design facilitates a comprehensive feature learning process, capturing intricate details and broader semantic contexts within medical images. We introduce a novel integration mechanism within the VMamba blocks to ensure seamless connectivity and information flow between the encoder and decoder paths, enhancing the segmentation performance. We conducted experiments on publicly available ACDC MRI Cardiac segmentation dataset, and Synapse CT Abdomen segmentation dataset. The results show that Mamba-UNet outperforms several types of UNet in medical image segmentation under the same hyper-parameter setting \footnote{The hyper-parameter setting includes: loss function, optimizer, training iterations, batch size, learning rate, same data splitting, etc.}. The source code and baseline implementations are available at \textcolor{mypink1}{https://github.com/ziyangwang007/Mamba-UNet}.

\end{abstract}
\section{Introduction}

Medical image segmentation is essential for diagnostics and treatments, and deep learning-based networks have shown dominant performance in this field \cite{long2015fully}. U-Net is one of the most essential architectures known for its symmetrical encoder-decoder style architecture and skip connections \cite{ronneberger2015u}, where various encoders and decoders extract feature information on different level, and skip connections enable the efficient transformation of feature information. Most of studies further explore U-Net with advanced network blocks techniques such as dense connections \cite{huang2017densely}, residual blocks \cite{he2016deep}, attention mechanisms \cite{woo2018cbam}, depthwise convolutions \cite{howard2017mobilenets}, and atrous convolutions \cite{yu2015multi,zhou2020acnn}, resulting in various modified UNet in CT, MRI, Ultrasound medical image segmentation \cite{oktay2018attention,ibtehaz2020multiresunet,li2018h,wang2021rar,zhang2020saunet,zhou2019unet++}.

\begin{figure*}[ht!]  
\centering  
\includegraphics[width=\linewidth]{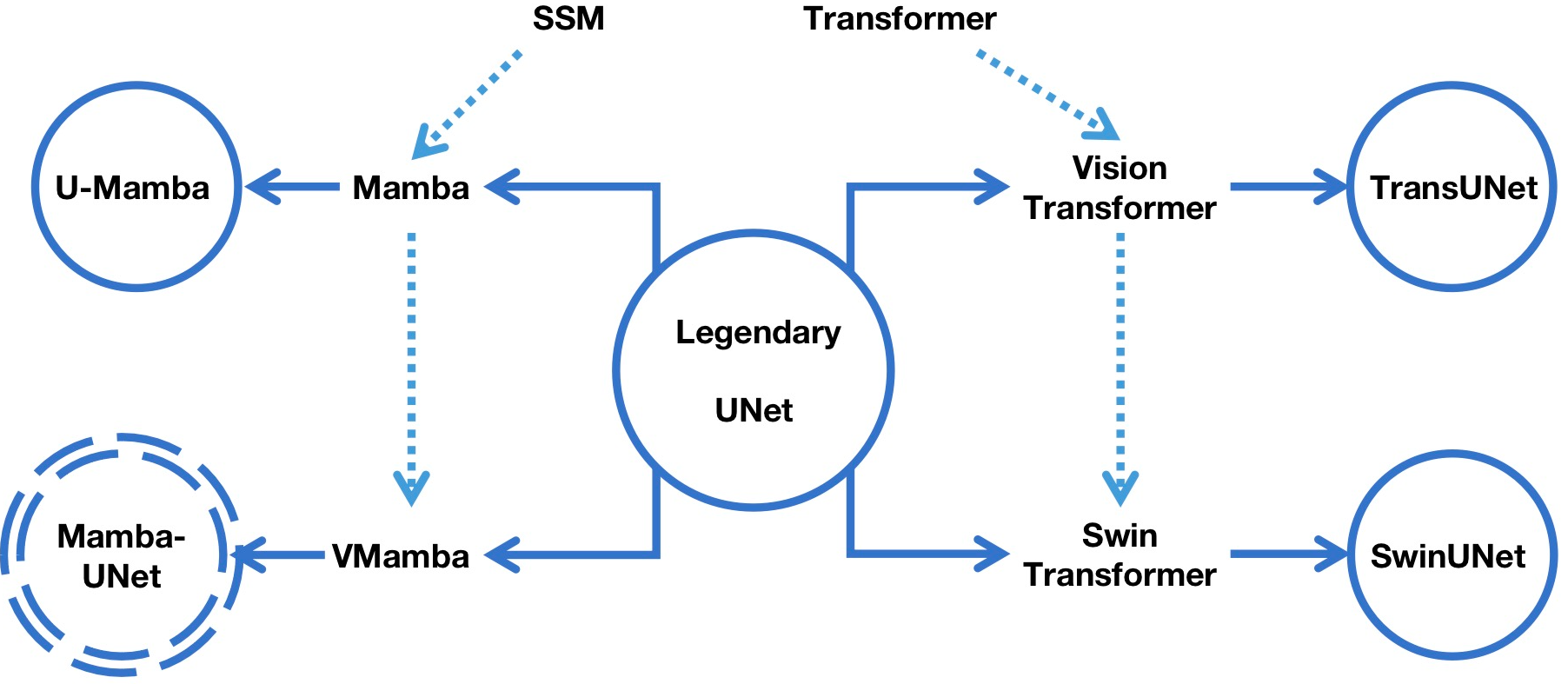}  
\caption{A brief introduction of the evolution of recent developments of UNet with incorporation of Transformer and State Space Models (SSM) for medical image segmentation.}  
\label{fig:intro}  
\end{figure*}

Motivated by the success of self-attention mechanisms from natural language processing \cite{vaswani2017attention}, ViT was the first to utilize a pure multi-head self-attention mechanism for the image recognition task with the state-of-the-art performance \cite{dosovitskiy2020image}. This showcase its promising capabilities in modeling long-range dependencies. Techniques like shift windows have further tailored ViT, resulting in Swin-Transformer \cite{liu2021swin}, which enhances their applicability in dense prediction tasks in computer vision, such as image segmentation, and detection \cite{liu2022video,xie2021self,liu2022swin}. In medical image segmentation, the integration of ViT with U-Net architectures, inspired by traditional CNN designs, has also led to various hybrid and pure ViT-based U-Nets. For instance, TransUNet is the first work to harness the feature learning power of ViT in the encoders of UNet \cite{chen2021transunet}. UNETR combines ViT with UNet for 3D segmentation \cite{hatamizadeh2022unetr}, while Swin-UNet and DCSUnet further explore purely Swin Vision Transformer network blocks with U-Net-based structure \cite{cao2022swin,wang2023densely}.

While Transformers excel in capturing long-range dependencies, their high computational cost, due to the quadratic scaling of the self-attention mechanism with input size, poses a challenge, particularly for high-resolution biomedical images \cite{xing2024segmamba,ma2024u}. Recent developments in State Space Models (SSMs) \cite{gu2023modeling,mehta2022long,wang2023selective}, especially Structured SSMs (S4) \cite{gu2021efficiently}, offer a promising solution with their efficient performance in processing long sequences. The Mamba model enhances S4 with a selective mechanism and hardware optimization, showing superior performance in dense data domains \cite{gu2023mamba}. The introduction of the Cross-Scan Module (CSM) in the Visual State Space Model (VMamba) further enhances Mamba's applicability to computer vision tasks by enabling the traversal of the spatial domain and converting non-causal visual images into ordered patch sequences \cite{liu2024vmamba}. Inspired by these capabilities, we propose leveraging Visual Mamba blocks (VSS) within the U-Net architecture to improve long-range dependency modeling in medical image analysis, resulting in Mamba-UNet. The evolution of U-Net with various network blocks and the positioning of our proposed Mamba-UNet are briefly illustrated in Figure \ref{fig:intro}.

\begin{figure*}[ht!]  
\centering  
\includegraphics[width=\linewidth]{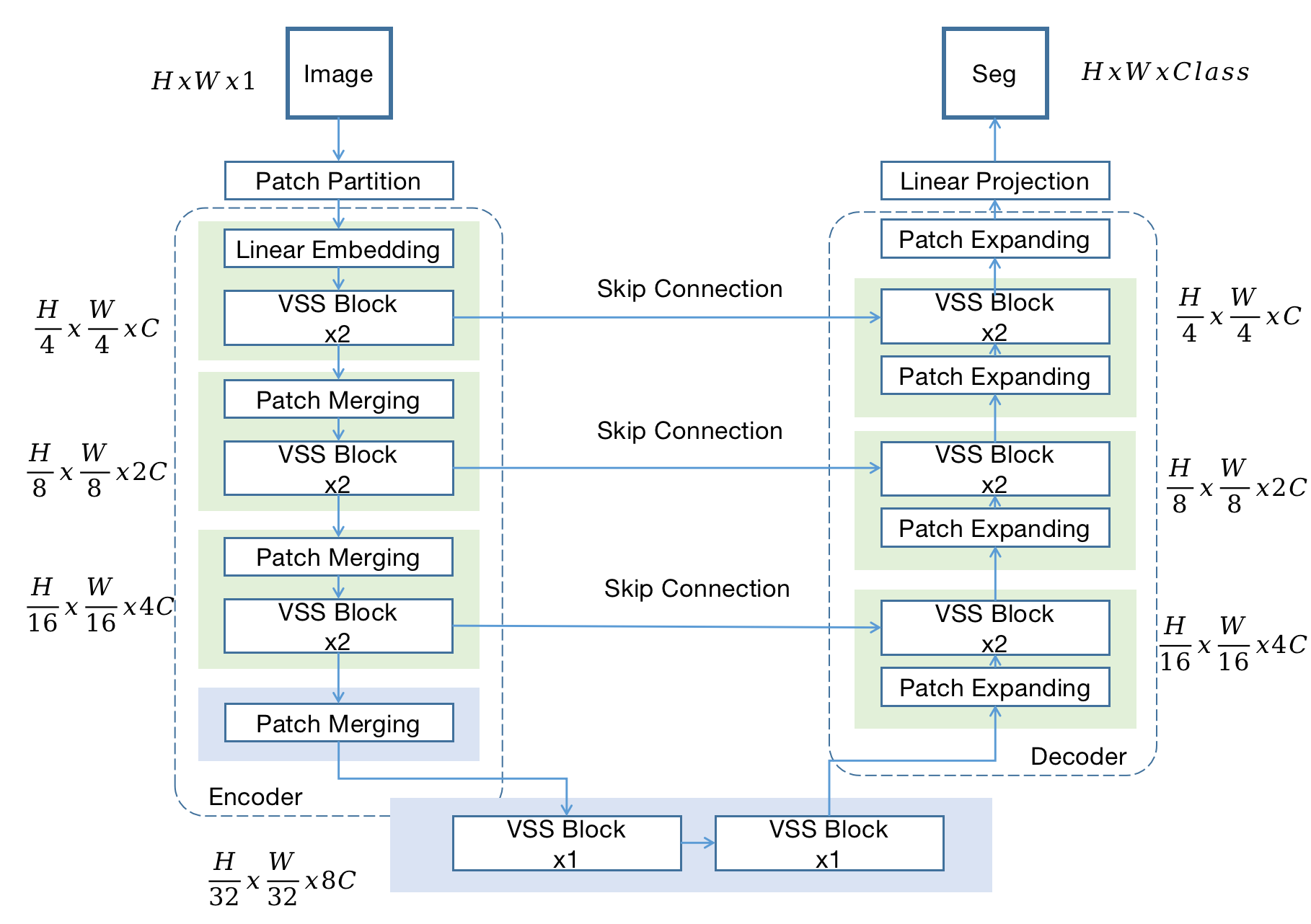}  
\caption{The architecture of Mamba-UNet, which is composed of encoder, bottleneck, decoder and skip connections. The encoder, bottleneck and decoder are all constructed based
on Visual Mamba block.}  
\label{fig:framework}  
\end{figure*}

\section{Approach}

\subsection{Architecture Overview}
The architecture of the proposed Mamba-UNet is sketched in Figure \ref{fig:framework}, which is motivated by UNet \cite{ronneberger2015u} and Swin-UNet \cite{cao2022swin}. The input 2D grey-scale image with the size of $H \times W \times 1$ is firstly spited into patch similar to ViT and VMamba \cite{dosovitskiy2020image,liu2024vmamba} then to 1-D sequence with the dimensions of $\frac{H}{4} \times \frac{W}{4} \times 16$. An initial linear embedding layer adjusts feature dimensions to an arbitrary size denoted as $C$. These patch tokens are then processed through multiple VSS blocks and patch merging layers, creating hierarchical features. Patch merging layers handle downsampling and dimension increase, while VSS blocks focus on learning feature representations. The output of each stage of encoder is with the resolution of $\frac{H}{4} \times \frac{W}{4} \times C$, $\frac{H}{8} \times \frac{W}{8} \times 2C$, $\frac{H}{16} \times \frac{W}{16} \times 4C$, and $\frac{H}{32} \times \frac{W}{32} \times 8C$, respectively. The decoder comprises VSS blocks and patch expanding layers following the encoder style enable the exact same feature size output, thus enhancing spatial details lost in downsampling through skip connections. In each of encoder and decoder, 2 VSS blocks are utilized, and the pretrained VMamba-Tiny \cite{liu2024vmamba} is loaded in the encoder, following the same process that Swin-UNet load the pretrained SwinViT-Tiny \cite{cao2022swin}. The details of VSS block, patch merging of encoder, and patch expanding of decoder is discussed in the following subsections.

\subsection{VSS Block}

\begin{figure*}[ht!]  
\centering  
\includegraphics[width=0.3\linewidth]{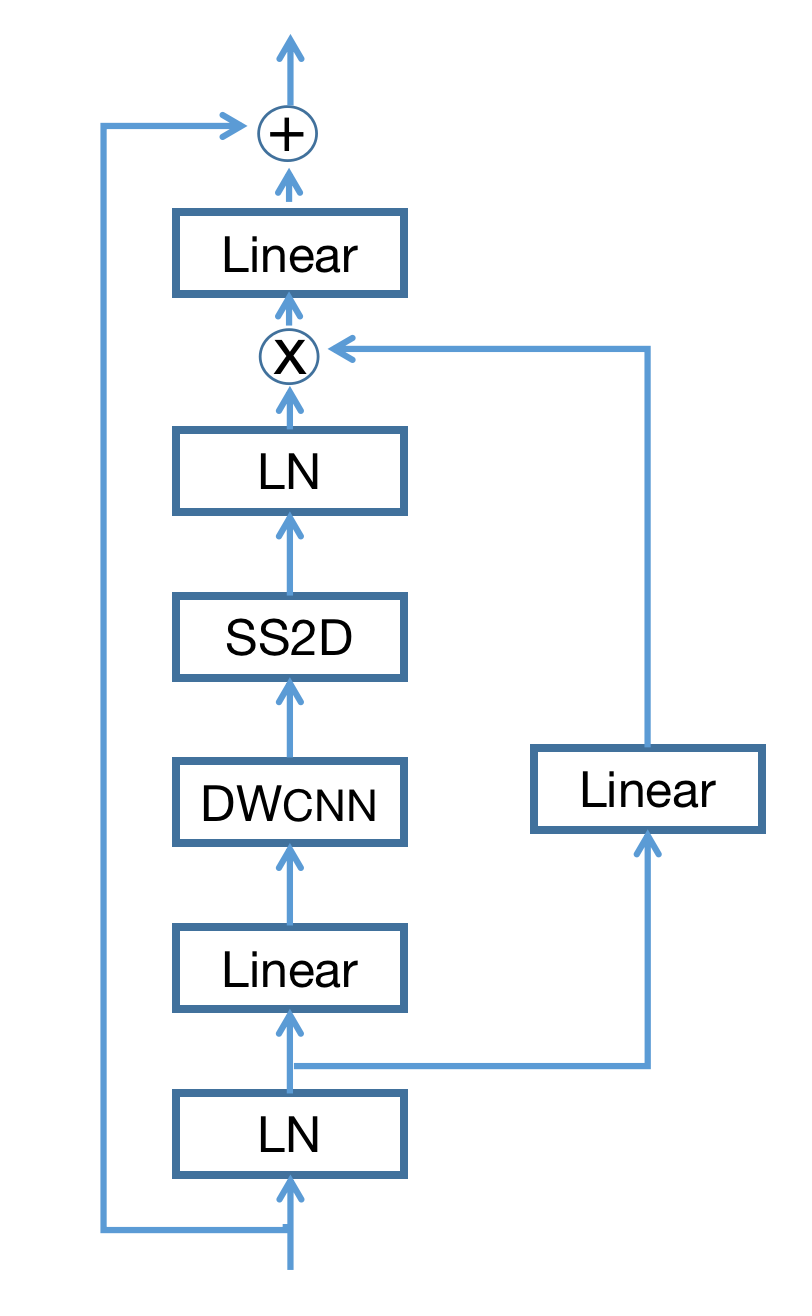}  
\caption{The detailed structure of the Visual State Space (VSS) Block.}  
\label{fig:VSS}  
\end{figure*}

The VSS network block is illustrated in Figure \ref{fig:VSS}, which is mainly based on Visual Mamba \cite{liu2024vmamba}.

Specifically, the conventional SSMs as a linear time-invariant system function to map $x(t) \in \mathbb{R} \mapsto y(t) \in \mathbb{R}$ through a hidden state $h(t) \in \mathbb{R}^N$, given $A \in \mathbb{C}^{N \times N}$ as the evolution parameter, $B, C \in \mathbb{C}^{N}$ as the projection parameters for a state size $N$, and skip connection $D \in \mathbb{C}^{1}$. The model can be formulated as linear ordinary differential equations (ODEs) in Eq \ref{ode},

\begin{equation}
\label{ode}
\begin{aligned}
h'(t) &= Ah(t) + Bx(t), \\
y(t) &= Ch(t) + Dx(t).
\end{aligned}
\end{equation}

The discrete version of this linear model can be transformed by zero-order hold given a timescale parameter $\Delta \in \mathbb{R}^{D}$.

\begin{equation}
\begin{aligned}
h_t &= \overline{A}h_{k-2} + \overline{B}x_k \\
y_t &= {C}h_k + \overline{D}x_k\\
\overline{A} &= e^{\Delta A} \\
\overline{B} &= (e^{\Delta A} - I) A^{-1}B \\
\overline{C} &= C 
\end{aligned}
\end{equation}
where $B,C \in \mathbb{R}^{D \times N}$. The approximation of $\overline{B}$ refined using first-order Taylor series $\overline{B} = \left(e^{\Delta A} - I\right) A^{-1} B \approx \left(\Delta A\right)\left(\Delta A\right)^{-1} \Delta B = \Delta B$. The Visual Mamba further introduce Cross-Scan Module (CSM) then integrate convolutional operations into the block, which is detailed in \cite{gu2023mamba,liu2024vmamba}. In the VSS block, the input feature first encounters a linear embedding layer, then bifurcates into dual pathways. One branch undergoes depth-wise convolution \cite{howard2017mobilenets} and SiLU activation \cite{shazeer2020glu}, proceeding to the SS2D module, and post-layer normalization, merges with the alternate stream post-SiLU activation. This VSS block eschews positional embedding, unlike typical vision transformers, opting for a streamlined structure sans the MLP phase, enabling a denser stack of blocks within the same depth budget. 

\subsection{Encoder}

In the encoder, $C$-dimensional tokenized inputs with reduced resolution undergo two consecutive VSS blocks for feature learning, maintaining dimension and resolution. The patch merging as downsampling process is utilized for three times in the encoder of Mamba-UNet, reduces the token count by $\frac{1}{2}$ while doubling feature dimensions by $2\times$, by segmenting inputs into quadrants by $\frac{1}{4}$, concatenating them, and then normalizing dimensions through a layernorm each time.

\subsection{Decoder}
Mirroring the encoder, the decoder utilizes two successive VSS blocks for feature reconstruction, employing patch expanding layers instead of merging ones for upscaling deep features \cite{cao2022swin}. These layers enhance resolution ($2 \times$ upscaling) while halving feature dimensions by $\frac{1}{2}$, exemplified by an initial layer that doubles feature dimensions before reorganizing and reducing them for resolution enhancement. 

\subsection{Bottleneck \& Skip Connetions}
Two VSS blocks are utilized for the bottleneck of Mamba-UNet. Each level of encoder and decoder employs skip connections to blend multi-scale features with upscaled outputs, enhancing spatial detail by merging shallow and deep layers. A subsequent linear layer maintains the dimensionality of this integrated feature set, ensuring consistency with the upscaled resolution.

\section{Experiments and Results}
\subsection{Data Sets}
{\bf Automated Cardiac Diagnosis Challenge} We conducted our experiments using the publicly available ACDC MRI cardiac segmentation dataset from the MICCAI 2017 Challenge \cite{bernard2018deep}. This dataset comprises MRI scans from 100 patients, annotated for multiple cardiac structures such as the right ventricle, and both the endocardial and epicardial walls of the left ventricle. It encompasses a diverse range of pathological conditions, categorized into five subgroups: normal, myocardial infarction, dilated cardiomyopathy, hypertrophic cardiomyopathy, and abnormal right ventricle, ensuring a broad distribution of feature information. 4 classes of Region of Interest (ROI) are validated in the ACDC. {\bf Synapse multi-organ segmentation Challenge} We also used the 30 abdominal CT scans in the MICCAI 2015 Multi-Atlas Abdomen Labeling Challenge, with 3779 axial contrast-enhanced abdominal clinical CT images in total \footnote{https://www.synapse.org/\#!Synapse:syn3193805/wiki/217789}. 9 classes of Region of Interest (ROI) are validated in the Synapse. To comply with the input requirements of the Swin-Transformer and Visual-Mamba pretrained network, all images were resized to $224{\times}224$. The dataset was partitioned such that 20\% of the images were allocated to the testing set, with the remainder used for training (including validation).

\subsection{Implementation Details}
The implementation was carried out on an Ubuntu 20.04 system, using Python 3.8.8, PyTorch 1.10, and CUDA 11.3. The hardware setup included an Nvidia GeForce RTX 3090 GPU and an Intel Core i9-10900K CPU. The average runtime was approximately 5 hours for ACDC dataset, and 12 hours for Synapse dataset, encompassing data transfer, model training, and inference processes. The dataset was specifically processed for 2D image segmentation. The Mamba-UNet model underwent training for 10,000 iterations with a batch size of 24. The Stochastic Gradient Descent (SGD) optimizer \cite{bottou-91c} was employed with a learning rate of 0.01, momentum of 0.9, and weight decay set to 0.0001. Network performance was evaluated on the validation set every 200 iterations, with model weights being saved only upon achieving a new best performance on the validation set.

\subsection{Baseline Methods}
 For comparative purposes, all baseline methods were also trained under identical hyperparameter configurations. The Mamba-UNet, along with other baseline methods including UNet \cite{ronneberger2015u}, Attention UNet \cite{oktay2018attention}, TransUNet \cite{chen2021transunet}, and Swin-UNet \cite{cao2022swin} are directly compared.

\subsection{Evaluation Metrics}
The assessment of Mamba-UNet against baseline methods utilizes a broad spectrum of evaluation metrics. Similarity measures, which are preferred to be higher, include: Dice, Intersection over Union (IoU), Accuracy, Precision, Sensitivity, and Specificity, denoted with an upward arrow ($\uparrow$) to indicate that higher values signify better performance. Conversely, difference measures such as the Hausdorff Distance (HD) 95\% and Average Surface Distance (ASD), marked with a downward arrow ($\downarrow$), are desirable when lower, indicating closer resemblance between the predicted and ground truth segmentations.

\begin{equation}
Dice = \frac{2 \times TP}{2 \times TP + FP + FN}
\end{equation}

\begin{equation}
\text{Accuracy} = \frac{TP + TN}{TP + TN + FP + FN}
\end{equation}

\begin{equation}
\text{Precision} = \frac{TP}{TP + FP}
\end{equation}

\begin{equation}
\text{Sensitivity} = \frac{TP}{TP + FN}
\end{equation}

\begin{equation}
\text{Specificity} = \frac{TN}{TN + FP}
\end{equation}

Where, \( TP \) represents the number of true positives, \( TN \) denotes the number of true negatives, \( FP \) signifies the number of false positives, and \( FN \) stands for the number of false negatives.

\begin{equation}
\text{Hausdorff Distance (HD) 95\%} = \max\left(\max_{a \in A} \min_{b \in B} d(a,b), \max_{b \in B} \min_{a \in A} d(a,b)\right)_{95\%}
\end{equation}

\begin{equation}
\text{Average Surface Distance (ASD)} = \frac{1}{|A|+|B|} \left(\sum_{a \in A} \min_{b \in B} d(a,b) + \sum_{b \in B} \min_{a \in A} d(a,b)\right)
\end{equation}

Where, $a$ and $b$ represent the sets of points on the predicted and ground truth surfaces, respectively. $d(a,b)$ denotes the Euclidean distance between two points. $95\%$ is a modified version of the Hausdorff Distance, focusing on the 95th percentile of the distances to reduce the impact of outliers.

\subsection{Qualitative Results}
\begin{figure*}[ht!]  
\centering  
\includegraphics[width=\linewidth]{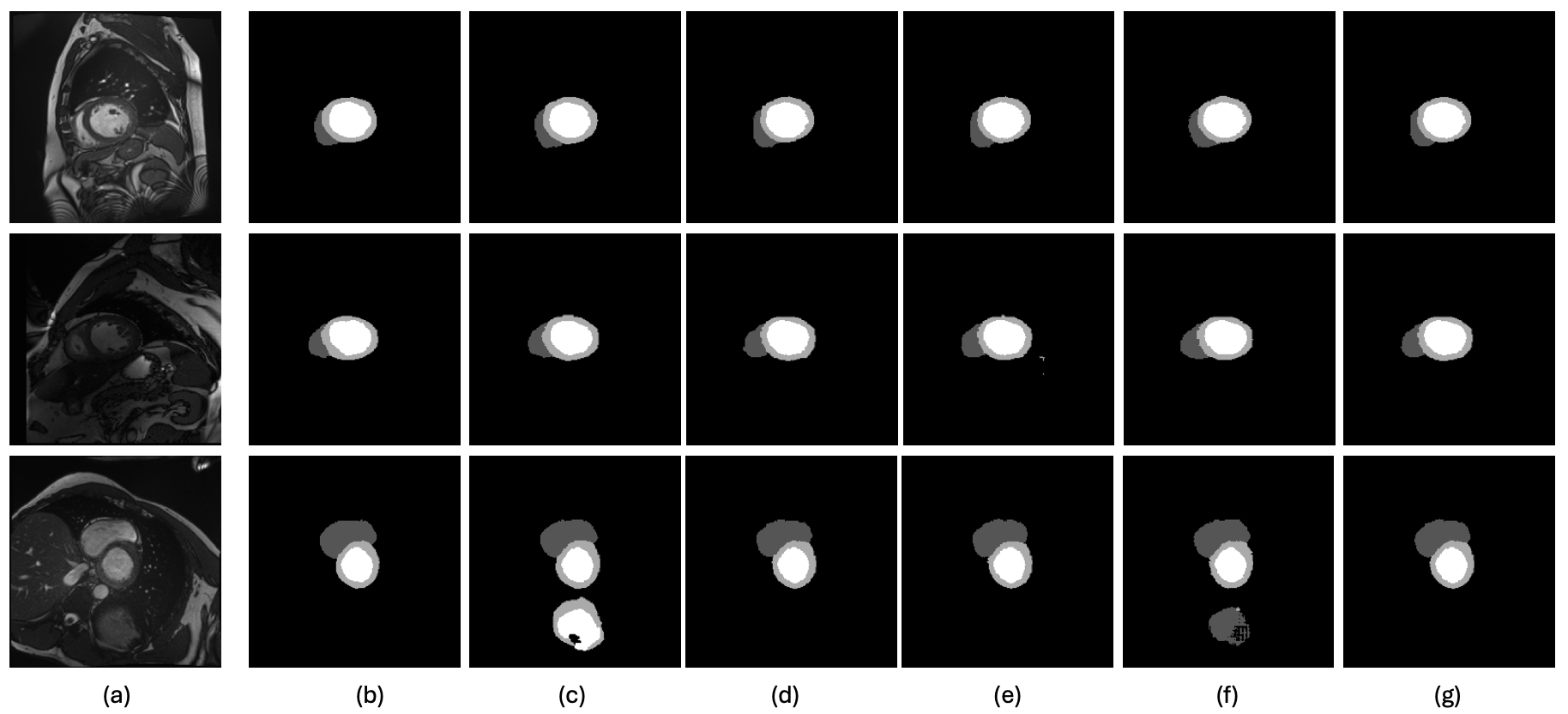}  
\caption{The visual comparison of segmentation results of Mamba-UNet and other segmentation methods against ground truth on ACDC MRI Cardiac Segmentation Dataset. (a) Raw MRI Image, (b) Ground Truth, (c)UNet, (d) Attention UNet, (e) TransUNet, (f) Swin-UNet, (g) Mamba-UNet.}  
\label{fig:comparsion}  
\end{figure*}

\begin{figure*}[ht!]  
\centering  
\includegraphics[width=\linewidth]{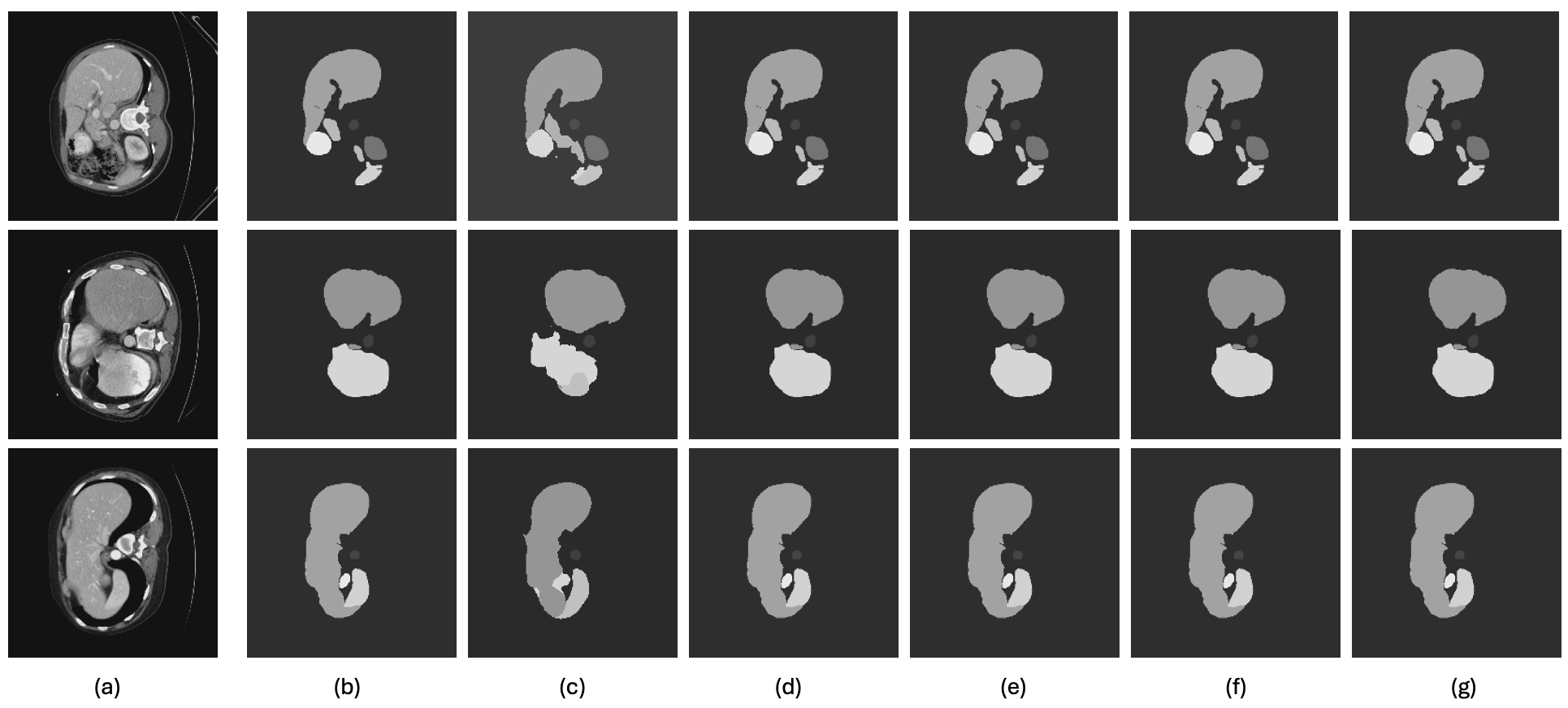}  
\caption{The visual comparison of segmentation results of Mamba-UNet and other segmentation methods against ground truth on Synapse CT Abdomen Segmentation Dataset. (a) Raw MRI Image, (b) Ground Truth, (c)UNet, (d) Attention UNet, (e) TransUNet, (f) Swin-UNet, (g) Mamba-UNet.}  
\label{fig:comparsionv2}  
\end{figure*}

Figure~\ref{fig:comparsion} and Figure~\ref{fig:comparsionv2} illustrates three randomly selected sample raw images, corresponding ground truth, and inference of all baseline methods including Mamba-UNet on ACDC and Synapse dataset, respectively.

\subsection{Quantitative Results} 
\begin{figure*}[ht!]  
\centering  
\includegraphics[width=\linewidth]{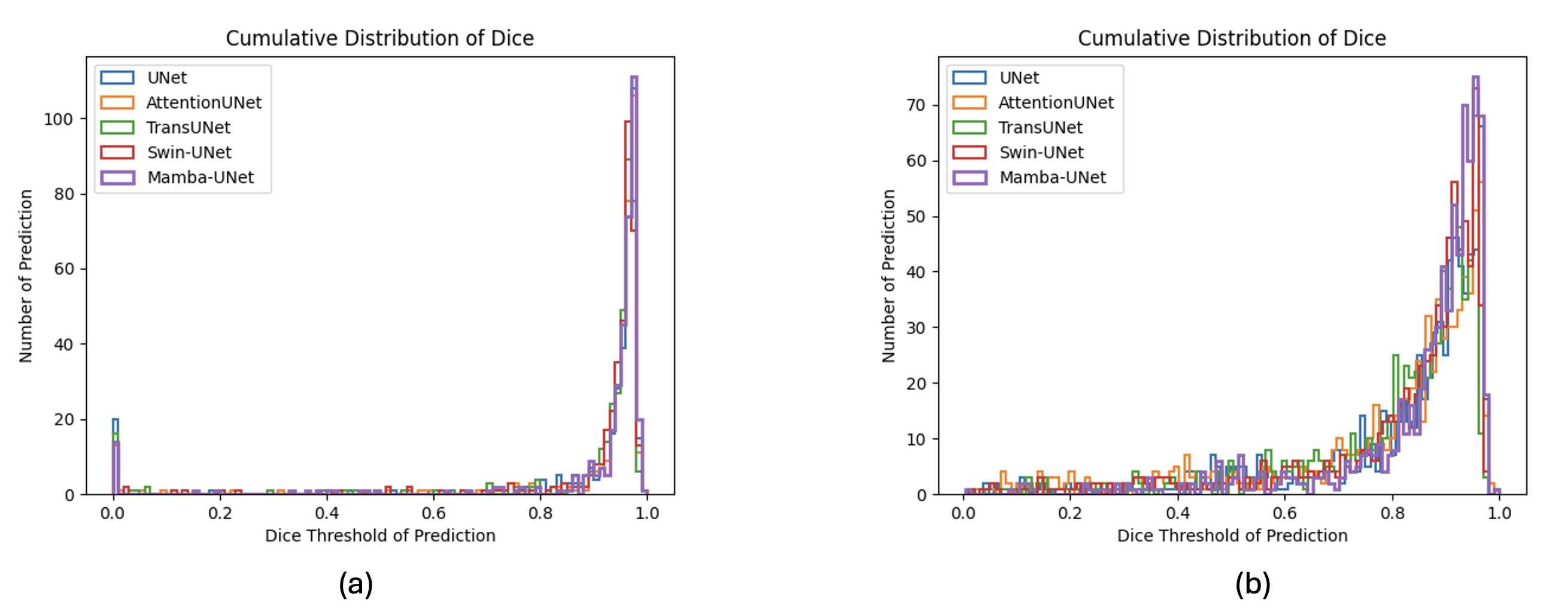}  
\caption{The histogram of the Dice distribution of Mamba-UNet and other segmentation methods against ground truth. (a) Test on ACDC MRI Cardiac Dataset. (b) Test on Synapse CT Abdomen Dataset.}  
\label{fig:histogram}  
\end{figure*}

Table \ref{tablebaseline} and Table \ref{tablebaselinev2} report the direct comparison of Mamba-UNet against other segmentation networks including similarity measures and difference measures on two dataset. The best performance is with {\bf Bold}, and the second best performance of Mamba-UNet is with \underline{Underline}. Quantitative results demonstrates that Mamba-UNet is morel likely to predict precise segmentation masks. To further validate the Mamba-UNet on test set, we also validate on the image by image fashion, and the distribution of segmentation prediction according to Dice-Coefficient is sketched in Figure \ref{fig:histogram}, where the X-axis is the Dice-Coefficient, and Y-axis is the number of predictions. This histogram further demonstrates that Mamba-UNet is more likely to provide prediction with high Dice-Coefficient performance.

\begin{table*}[htbp]
\caption{Direct Comparison of Segmentation Networks Performance on ACDC MRI Cardiac Test Set}
\centering
\begin{tabular}{c|cccccc|cc}
\hline
Framework & Dice$\uparrow$ & IoU$\uparrow$ & Acc$\uparrow$ & Pre$\uparrow$ & Sen$\uparrow$ & Spe$\uparrow$ & HD$\downarrow$ & ASD$\downarrow$  \\
\hline
UNet  & 0.9248 & 0.8645&0.9969 & 0.9157 & {\bf 0.9364} & {\bf 0.9883} & 2.7655 & 0.8180 \\
AttentionUNet &0.9249 & 0.8647&0.9970 & 0.9239 & 0.9260 & 0.9858 &3.4156 & 0.9765   \\
TransUNet & 0.9196 & 0.8561&0.9968 & 0.9187 & 0.9207 & 0.9846 & 2.7742 & 0.8324 \\
Swin-UNet  & 0.9188 & 0.8545&0.9968 & 0.9151 & 0.9231 & 0.9857 & 3.1817 & 0.9932\\
\hline
Mamba-UNet & {\bf 0.9281} & {\bf 0.8698} & {\bf 0.9972} & {\bf 0.9275} & \underline{0.9289} & \underline{0.9859} & {\bf 2.4645} & {\bf 0.7677} \\
\hline
\end{tabular}
\label{tablebaseline}
\end{table*}

\begin{table*}[htbp]
\caption{Direct Comparison of Segmentation Networks Performance on Synapse CT abdominal Test Set}
\centering
\begin{tabular}{c|cccccc|cc}
\hline
Framework & Dice$\uparrow$ & IoU$\uparrow$ & Acc$\uparrow$ & Pre$\uparrow$ & Sen$\uparrow$ & Spe$\uparrow$ & HD$\downarrow$ & ASD$\downarrow$  \\
\hline
UNet & 0.6299 & 0.5198 & 0.9969 & 0.6224 & 0.6541 & 0.9889 & 37.7342 & 10.0725 \\
AttentionUNet & 0.6069 & 0.4964 & 0.9962 & 0.5959 & 0.6394 & {\bf 0.9890} & 74.2449 & 25.8229 \\
TransUNet & 0.6092 & 0.5017 & 0.9970 & 0.6027 & 0.6318 & 0.9858 & 33.8126 & 10.8979 \\
Swin-UNet  & 0.6178 & 0.5121 & 0.9972 & 0.6126 & 0.6357 & 0.9859 & 30.5414 & 9.2504 \\
\hline
Mamba-UNet & {\bf 0.6429} & {\bf 0.5405} & {\bf 0.9975} & {\bf 0.6452} & {\bf 0.6603} & {\bf 0.9890} & {\bf 24.4725} & {\bf 6.4717} \\
\hline
\end{tabular}
\label{tablebaselinev2}
\end{table*}

\section{Conclusion}
In this paper, we introduced Mamba-UNet, which is a purely Visual Mamba block-based UNet style network for medical image segmentation. The performance demonstrates that Mamba-UNet superior performance against classical similar network such as UNet and Swin-UNet. In the future, we aim to conduct more in-depth explorations on more medical image segmentation tasks from different modalities and targets, with comparisons to more segmentation backbones. Besides, we aim to extend Mamba-UNet to 3D medical images, and semi/weakly-supervised learning \cite{SemiSurvey} to further enhance the developments in medical imaging.

\bibliographystyle{splncs04}
\bibliography{ref}

\begin{thebibliography}{10}
\providecommand{\url}[1]{\texttt{#1}}
\providecommand{\urlprefix}{URL }
\providecommand{\doi}[1]{https://doi.org/#1}

\bibitem{bernard2018deep}
Bernard, O., et~al.: Deep learning techniques for automatic mri cardiac multi-structures segmentation and diagnosis: is the problem solved? IEEE transactions on medical imaging  \textbf{37}(11),  2514--2525 (2018)

\bibitem{bottou-91c}
Bottou, L.: Stochastic gradient learning in neural networks. In: Proceedings of Neuro-N\^imes 91. EC2, Nimes, France (1991)

\bibitem{cao2022swin}
Cao, H., Wang, Y., Chen, J., Jiang, D., Zhang, X., Tian, Q., Wang, M.: Swin-unet: Unet-like pure transformer for medical image segmentation. In: European conference on computer vision. pp. 205--218. Springer (2022)

\bibitem{chen2021transunet}
Chen, J., Lu, Y., Yu, Q., Luo, X., Adeli, E., Wang, Y., Lu, L., Yuille, A.L., Zhou, Y.: Transunet: Transformers make strong encoders for medical image segmentation. arXiv preprint arXiv:2102.04306  (2021)

\bibitem{dosovitskiy2020image}
Dosovitskiy, A., Beyer, L., Kolesnikov, A., Weissenborn, D., Zhai, X., Unterthiner, T., Dehghani, M., Minderer, M., Heigold, G., Gelly, S., et~al.: An image is worth 16x16 words: Transformers for image recognition at scale. arXiv preprint arXiv:2010.11929  (2020)

\bibitem{gu2023modeling}
Gu, A.: Modeling Sequences with Structured State Spaces. Ph.D. thesis (2023)

\bibitem{gu2023mamba}
Gu, A., Dao, T.: Mamba: Linear-time sequence modeling with selective state spaces. arXiv preprint arXiv:2312.00752  (2023)

\bibitem{gu2021efficiently}
Gu, A., Goel, K., R{\'e}, C.: Efficiently modeling long sequences with structured state spaces. arXiv preprint arXiv:2111.00396  (2021)

\bibitem{hatamizadeh2022unetr}
Hatamizadeh, A., Tang, Y., Nath, V., Yang, D., Myronenko, A., Landman, B., Roth, H.R., Xu, D.: Unetr: Transformers for 3d medical image segmentation. In: Proceedings of the IEEE/CVF winter conference on applications of computer vision. pp. 574--584 (2022)

\bibitem{he2016deep}
He, K., Zhang, X., Ren, S., Sun, J.: Deep residual learning for image recognition. In: Proceedings of the IEEE conference on computer vision and pattern recognition. pp. 770--778 (2016)

\bibitem{howard2017mobilenets}
Howard, A.G., Zhu, M., Chen, B., Kalenichenko, D., Wang, W., Weyand, T., Andreetto, M., Adam, H.: Mobilenets: Efficient convolutional neural networks for mobile vision applications. arXiv preprint arXiv:1704.04861  (2017)

\bibitem{huang2017densely}
Huang, G., Liu, Z., Van Der~Maaten, L., Weinberger, K.Q.: Densely connected convolutional networks. In: Proceedings of the IEEE conference on computer vision and pattern recognition. pp. 4700--4708 (2017)

\bibitem{ibtehaz2020multiresunet}
Ibtehaz, N., Rahman, M.S.: Multiresunet: Rethinking the u-net architecture for multimodal biomedical image segmentation. Neural networks  \textbf{121},  74--87 (2020)

\bibitem{SemiSurvey}
Jiao, R., Zhang, Y., Ding, L., Xue, B., Zhang, J., Cai, R., Jin, C.: Learning with limited annotations: A survey on deep semi-supervised learning for medical image segmentation. Computers in Biology and Medicine  (2023)

\bibitem{li2018h}
Li, X., Chen, H., Qi, X., Dou, Q., Fu, C.W., Heng, P.A.: H-denseunet: hybrid densely connected unet for liver and tumor segmentation from ct volumes. IEEE transactions on medical imaging  \textbf{37}(12),  2663--2674 (2018)

\bibitem{liu2024vmamba}
Liu, Y., Tian, Y., Zhao, Y., Yu, H., Xie, L., Wang, Y., Ye, Q., Liu, Y.: Vmamba: Visual state space model. arXiv preprint arXiv:2401.10166  (2024)

\bibitem{liu2022swin}
Liu, Z., Hu, H., Lin, Y., Yao, Z., Xie, Z., Wei, Y., Ning, J., Cao, Y., Zhang, Z., Dong, L., et~al.: Swin transformer v2: Scaling up capacity and resolution. In: Proceedings of the IEEE/CVF conference on computer vision and pattern recognition. pp. 12009--12019 (2022)

\bibitem{liu2021swin}
Liu, Z., Lin, Y., Cao, Y., Hu, H., Wei, Y., Zhang, Z., Lin, S., Guo, B.: Swin transformer: Hierarchical vision transformer using shifted windows. In: Proceedings of the IEEE/CVF international conference on computer vision. pp. 10012--10022 (2021)

\bibitem{liu2022video}
Liu, Z., Ning, J., Cao, Y., Wei, Y., Zhang, Z., Lin, S., Hu, H.: Video swin transformer. In: Proceedings of the IEEE/CVF conference on computer vision and pattern recognition. pp. 3202--3211 (2022)

\bibitem{long2015fully}
Long, J., Shelhamer, E., Darrell, T.: Fully convolutional networks for semantic segmentation. In: Proceedings of the IEEE conference on computer vision and pattern recognition. pp. 3431--3440 (2015)

\bibitem{ma2024u}
Ma, J., Li, F., Wang, B.: U-mamba: Enhancing long-range dependency for biomedical image segmentation. arXiv preprint arXiv:2401.04722  (2024)

\bibitem{mehta2022long}
Mehta, H., Gupta, A., Cutkosky, A., Neyshabur, B.: Long range language modeling via gated state spaces. arXiv preprint arXiv:2206.13947  (2022)

\bibitem{oktay2018attention}
Oktay, O., Schlemper, J., Folgoc, L.L., Lee, M., Heinrich, M., Misawa, K., Mori, K., McDonagh, S., Hammerla, N.Y., Kainz, B., et~al.: Attention u-net: Learning where to look for the pancreas. arXiv preprint arXiv:1804.03999  (2018)

\bibitem{ronneberger2015u}
Ronneberger, O., et~al: U-net: Convolutional networks for biomedical image segmentation. In: MICCAI (2015)

\bibitem{shazeer2020glu}
Shazeer, N.: Glu variants improve transformer. arXiv preprint arXiv:2002.05202  (2020)

\bibitem{vaswani2017attention}
Vaswani, A., Shazeer, N., Parmar, N., Uszkoreit, J., Jones, L., Gomez, A.N., Kaiser, {\L}., Polosukhin, I.: Attention is all you need. Advances in neural information processing systems  \textbf{30} (2017)

\bibitem{wang2023selective}
Wang, J., Zhu, W., Wang, P., Yu, X., Liu, L., Omar, M., Hamid, R.: Selective structured state-spaces for long-form video understanding. In: Proceedings of the IEEE/CVF Conference on Computer Vision and Pattern Recognition. pp. 6387--6397 (2023)

\bibitem{wang2023densely}
Wang, Z., Su, M., Zheng, J.Q., Liu, Y.: Densely connected swin-unet for multiscale information aggregation in medical image segmentation. In: 2023 IEEE International Conference on Image Processing (ICIP). pp. 940--944. IEEE (2023)

\bibitem{wang2021rar}
Wang, Z., Zhang, Z., Voiculescu, I.: Rar-u-net: a residual encoder to attention decoder by residual connections framework for spine segmentation under noisy labels. In: 2021 IEEE International Conference on Image Processing (ICIP). pp. 21--25. IEEE (2021)

\bibitem{woo2018cbam}
Woo, S., Park, J., Lee, J.Y., Kweon, I.S.: Cbam: Convolutional block attention module. In: Proceedings of the European conference on computer vision (ECCV). pp. 3--19 (2018)

\bibitem{xie2021self}
Xie, Z., Lin, Y., Yao, Z., Zhang, Z., Dai, Q., Cao, Y., Hu, H.: Self-supervised learning with swin transformers. arXiv preprint arXiv:2105.04553  (2021)

\bibitem{xing2024segmamba}
Xing, Z., Ye, T., Yang, Y., Liu, G., Zhu, L.: Segmamba: Long-range sequential modeling mamba for 3d medical image segmentation. arXiv preprint arXiv:2401.13560  (2024)

\bibitem{yu2015multi}
Yu, F., Koltun, V.: Multi-scale context aggregation by dilated convolutions. arXiv preprint arXiv:1511.07122  (2015)

\bibitem{zhang2020saunet}
Zhang, Y., Yuan, L., Wang, Y., Zhang, J.: Sau-net: efficient 3d spine mri segmentation using inter-slice attention. In: Medical Imaging With Deep Learning. pp. 903--913. PMLR (2020)

\bibitem{zhou2020acnn}
Zhou, X.Y., Zheng, J.Q., Li, P., Yang, G.Z.: Acnn: a full resolution dcnn for medical image segmentation. In: 2020 IEEE International Conference on Robotics and Automation (ICRA). pp. 8455--8461. IEEE (2020)

\bibitem{zhou2019unet++}
Zhou, Z., Siddiquee, M.M.R., Tajbakhsh, N., Liang, J.: Unet++: Redesigning skip connections to exploit multiscale features in image segmentation. IEEE transactions on medical imaging  \textbf{39}(6),  1856--1867 (2019)

\end{thebibliography}

\end{document}